\documentclass[11pt]{article}
\usepackage{graphicx}
\usepackage{amsmath,amssymb}
\newtheorem{theorem}{Theorem}

\def\1{{\bf 1}}

\def\be{\begin{equation}}
\def\ee{\end{equation}}

\begin{document}
\noindent {\tiny J. Theor. Biol. 284(1) 149-157 (2011), PMID: 21723301,  DOI:10.1016/j.jtbi.2011.06.024, submitted 11/03/2011, revised 15/06/2011, accepted 17/06/2011, published online 28/06/2011, http://authors.elsevier.com/offprints/YJTBI6526/174fceebd1ad042c46f17745c5be5f47}\\ 
{\bf--------------------------------------------------------------------------------------------------- }\\

\bigskip
	
\noindent {\large \bf A novel canonical dual computational approach for prion AGAAAAGA amyloid fibril molecular modeling}\\

\bigskip

\noindent {\Large Jiapu Zhang$^*$, David Y Gao, John Yearwood}\\

\noindent Centre in Informatics and Applied Optimization\& Graduate School of Sciences, Informatics Technology and Engineering,  University of Ballarat, Mount Helen, VIC 3353, Australia.\\
$^*$Emails: j.zhang@ballarat.edu.au, jiapu\_zhang@hotmail.com, Tel: 61-4 2348 7360, 61-3-5327 9809\\

\noindent $^*$Corresponding author\\

\noindent {\bf Abstract}\\ 
Many experimental studies have shown that the prion AGAAAAGA palindrome hydrophobic region (113-120) has amyloid fibril forming properties and plays an important role in prion diseases. However, due to the unstable, noncrystalline and insoluble nature of the amyloid fibril, to date structural information on AGAAAAGA region (113-120) has been very limited. This region falls just within the N-terminal unstructured region PrP (1-123) of prion proteins. Traditional X-ray crystallography and nuclear magnetic resonance (NMR) spectroscopy experimental methods cannot be used to get its structural information. Under this background, this paper introduces a novel approach of the canonical dual theory to address the 3D atomic-resolution structure of prion AGAAAAGA amyloid fibrils. The novel and powerful canonical dual computational approach introduced in this paper is for the molecular modeling of prion AGAAAAGA amyloid fibrils, and that the optimal atomic-resolution structures of prion AGAAAAGA amyloid fibils presented in this paper are useful for the drive to find treatments for prion diseases in the field of medicinal chemistry. Overall, this paper presents an important method and provides useful information for treatments of prion diseases. Overall, this paper could be of interest to the general readership of Journal of Theoretical Biology.\\

\noindent {\bf Highlights}\\
$\blacktriangleright$ Study of prion AGAAAAGA amyloid fibril molecular structures. $\blacktriangleright$ Sum of van der Waals radii regarding the minimization point of the Lennard-Jones potential energy. $\blacktriangleright$ Mathematical model into a global optimization molecular distance geometry problem. $\blacktriangleright$ Use of a novel canonical dual computational approach to solve the model.  $\blacktriangleright$ Use of computational chemistry Amber package to refine the model.\\ 


\section{Introduction}
According to a recent comprehensive review (Chou, 2011), to develop a useful model for biological systems, the following things were usually needed to consider: (i) the material of benchmark used to develop and test the model, (ii) the formulation of modeling method, (iii) operating procedures during the modeling process, (iv) properly perform the cross-validation tests to objectively evaluate the anticipated accuracy of the model, and (v) web-server establishment. Below, let us elaborate some of these procedures. In this paper, the material used to develop the model is 3NHC.pdb and its 3D-crystal structures; the modeling method is the Mathematical Optimization methods of the canonical dual theory (CDT) (Gao, 2000, Gao et al., 2010, Gao et al., 2012) ({\it procedure 1}) and of the Amber 11 package's steepest Descent (SD) method (Case et al., 2010) and Conjugate Gradient (CG) method (Case et al., 2010, Sun et al., 2001) ({\it procedure 2}); and the test to the accuracy of the model is performed by the RMSD (root-mean-square deviation) value of last snapshots between {\it procedures 1--2}.\\

Various computational molecular dynamics approaches have been used to study PrP (106–-126) (Kuwata et al., 2003, Wagoner 2010) but, to the best of our knowledge, to predict molecular structures of prion AGAAAAGA amyloid fibrils the computational approaches are few (Zhang, 2011, Zhang et al., 2011). Zhang (2011) successfully constructed three AGAAAAGA amyloid fibril models by the standard simulated annealing (SA) method and several traditional optimization methods within AMBER 10 package. In (Zhang et al., 2011), the hybrid simulated annealing discrete gradient (SADG) method was successfully used for modeling two AGAAAAGA amyloid fibril models (instead of the Insight II (http://accelrys.com) package used in (Zhang, 2011)), and then the models were refined/optimized by the SDCG methods, SA method and SDCG methods again. In this paper, all the optimization approaches of (Zhang, 2011, Zhang et al., 2011) will be replaced by the optimization theory of CDT. Numerical computational results show that the optimization approaches of CDT have a very perfect performance. It is even no need to do furthermore SDCG refinements by the AMBER package. We could not do comparisons (for example, the angstrom values between adjacent $\beta$-sheets and $\beta$-strands) for the models of (Zhang, 2011, Zhang et al., 2011) and of this paper, because these models have different number of chains and different structural Classes listed in (Kuwata et al., 2003).\\

As we all know, the disease prions PrP$^{Sc}$ are rich in $\beta$-sheets amyloid fibrils (about 43\% $\beta$-sheet) (Griffith, 1967). There are some classical works on the $\beta$-sheets and $\beta$-barrels (Chou et al., 1983a, 1990a, b, 1991). %\cite{chouns1983a, choucm1990a, chouns1990b, chouc1991}.\\ 
X-ray crystallography and nuclear magnetic resonance (NMR) spectroscopy are two powerful tools to determine the protein 3D structure. However, not all proteins can be successfully crystallized, particularly for membrane proteins. Although NMR is indeed a very powerful tool in determining the 3D structures of membrane proteins (see, e.g., (Schnell et al., 2008, Oxenoid et al., 2005, Call et al., 2010, Pielak et al., 2010, Pielak et al., 2009,Wang et al., 2009) and a recent review (Pielak et al., 2011)), it is also time-consuming and costly. To acquire the structural information in a timely manner, one has to resort to various structural bioinformatics tools (see, e.g., (Chou, 2004b, Chou, 2004c, Chou, 2004d, Chou, 2005b) and a comprehensive review (Chou, 2004c)). Particularly, computational approaches allow us to obtain a description of the protein 3D structure at a submicroscopic level. Under many circumstances, due to the unstable, noncrystalline and insoluble nature of the amyloid fibrils, it is very difficult to use traditional X-ray and NMR experimental methods to obtain atomic-resolution structures of amyloid fibrils (Tsai, 2005, Zheng et al., 2006). Although X-ray and NMR techniques cannot determine the 3D structures of some proteins and their binding interactions with ligands in a timely manner that are important for drug design and basic research, many structural bioinformatics tools can play a complementary role in this regard as demonstrated by a series papers published recently (see, e.g. (Cai et al., 2011, Chou, 2004a, Chou, 2005a, Chou et al., 2003, Du et al., 2007, Du et al., 2010, Gong et al., 2009, Liao et al., 2011, Wang et al., 2010, Wang et al., 2009, Wei et al., 2009)). This paper, in some sense, presents a structural bioinformatics tool in view of the CDT-based mathematical optimization theory.\\

The accuracy of the models presented in this paper is tested by the RMSD value. The last snapshot of {\it procedure 2} will be superposed onto the last snapshot of {\it procedure 1}, and the RMSD value is zero after the alignment by VMD 1.8.7beta5 (Humphrey et al., 1996). This implies to us that the CDT strategy can accurately built the prion AGAAAAGA amyloid fibril models. To test the accuracy of their model, some examination validation methods are always used. In developing a prediction model or algorithm, the following three cross-validation methods are often used for examining its effectiveness in practical application: independent dataset test, subsampling (5-fold or 10-fold cross-validation) test, and jackknife test (Chou et al., 1995). However, as demonstrated by Eqs.28-32 of (Chou, 2011), among the three cross-validation methods, the jackknife test is deemed the least arbitrary that can always yield a unique result for a given benchmark dataset, and hence has been increasingly used and widely recognized by investigators to examine the accuracy of various models and predictors (see, e.g. (Chen et al., 2009, Chou et al., 2011, Ding et al., 2009, Hayat et al., 2011, Kandaswamy et al., 2011, Lin et al., 2011,Mohabatkar, 2010, Zeng et al., 2009, Zhou et al., 2007); all these papers reflect the current trend of increasingly and widely using the jackknife test to examine varieties of models or predictors).\\

There is another criteria to evaluate the models. To avoid homology bias and remove the redundant sequences from the benchmark dataset, a cutoff threshold of 25\% was recommended (Chou, 2011, Chou et al., 2011) to exclude those proteins from the benchmark datasets that have equal to or greater than 25\% sequence identity to any other. However, in this study we did not use such a stringent criterion because the currently available data do not allow us to do so. Otherwise, the numbers of proteins left would be too few to have statistical significance.\\% As it is well known, the more stringent of a benchmark dataset in excluding homologous sequences, the more reliable the results derived based on their model. According to a recent comprehensive review (Chou, 2011), an ideal cutoff threshold should be set at 25%, meaning that none of protein sequences included in the benchmark dataset has greater than 25% pairwise sequence identity to any other.\\ 

The last procedure to develop a useful model for biological systems is a web-server establishment. Since user-friendly and publicly accessible web-servers represent the future direction for developing practically more useful models, simulated methods, or predictors (Chou et al., 2009), we shall make efforts in our future work to provide a web-server for the method presented in this paper.\\

This paper addresses an important problem on neurodegenerative amyloid fibril or plaque diseases. The rest of this paper is arranged as follows. In the next section, i.e. Section 2, the CDT will be briefly introduced and its effectiveness will be illuminated by applying the CDT-based optimization approach to a well-known system of minimizing the Double Well Potential function. In Section 3, the molecular modeling works of prion AGAAAAGA amyloid fibrils will be done. Section 3 also successfully gains the optimal prion AGAAAAGA amyloid fibril models by the applications of the CDT-based optimization theory. Furthermore refinement/optimization to these models by the SDCG methods of the package Amber 11 will also be done in Section 3. The zero RMSD value implies to us that the CDT optimization strategy can accurately obtain the prion AGAAAAGA amyloid fibril models. Thus, when using the time-consuming and costly X-ray crystallography or NMR spectroscopy we still cannot determine the protein 3D structure, we may introduce computational approaches or novel mathematical formulations and physical concepts into molecular biology to study molecular structures. This concluding remark will be made in the last section, i.e. Section 4.\\    

\section{The Canonical Dual Approach}
We briefly introduce the CDT of (Gao et al., 2010, Gao et al., 2012, Gao, 2000) specially for solving the following minimization problem of the sum of fourth-order polynomials:
\begin{eqnarray}
&&(\mathcal{P}): \min_x \left\{ P(x) =\sum_{i=1}^m W_i (x)  + \frac{1}{2}x^TQx - x^Tf: x\in \mathbb{R}^n \right\} , \label{primeP}\\
&&where \quad W_i (x)= \frac{1}{2} \alpha_i \left( \frac{1}{2} x^TA_ix +b_i^Tx +c_i \right)^2, A_i=A^T_i, Q=Q^T\in \mathbb{R}^{n\times n}, \nonumber\\ 
&&b_i, f\in \mathbb{R}^n, c_i, \alpha_i \in \mathbb{R}^1, i=1,2,\dots, m, x\in \mathcal{X} \subset \mathbb{R}^n. \nonumber
\end{eqnarray}
The dual problem of $(\mathcal{P})$ is
\begin{eqnarray}
&&(\mathcal{P}^d): \max_{\varsigma} \left\{ P^d(\varsigma ) =\sum_{i=1}^m \left( c_i \varsigma_i -\frac{1}{2} \alpha_i^{-1} \varsigma^2 \right) -\frac{1}{2} F^T(\varsigma ) G^+(\varsigma ) F(\varsigma ): \varsigma \in S_a \right\} ,\label{dualPd}\\
&&where \quad F(\varsigma )=f-\sum_{i=1}^m \varsigma_i b_i, G(\varsigma )=Q+\sum_{i=1}^m \varsigma_i A_i, S_a=\{ \varsigma \in \mathbb{R}^m | F(\varsigma ) \in Col(G(\varsigma )) \}, \nonumber
\end{eqnarray}
$G^+$ denotes the Moore-Penrose generalized inverse of $G$, and $Col(G(\varsigma ))$ is the column space of $G(\varsigma )$. The prime-dual Gao-Strang complementary function of CDT (Gao et al., 2010, Gao et al., 2012, Gao, 2000) is
\begin{eqnarray}
\Xi (x,\varsigma )=\sum_{i=1}^m \left[ \left( \frac{1}{2} x^TA_ix +b_i^Tx +c_i \right) \varsigma_i -\frac{1}{2} \alpha_i^{-1} \varsigma_i^2 \right] +\frac{1}{2} x^TQx -x^Tf. \label{Gao-StrangComplementary} 
\end{eqnarray}
For $(\mathcal{P})$ and $(\mathcal{P}^d)$ we have the following CDT:
\begin{theorem} 
(Gao et al., 2010, Gao et al., 2012, Gao, 2000) The problem $(\mathcal{P}^d )$ is canonically dual to $(\mathcal{P} )$ in the sense that if $\bar{\varsigma}$ is a critical point of $P^d(\varsigma )$, then $\bar{x} =G^+(\bar{\varsigma} )F(\bar{\varsigma} )$ is a critical point of $P(x)$ on $\mathbb{R}^n$, and $P(\bar{x} )=P^d (\bar{\varsigma} )$. Moreover, if $\bar{\varsigma} \in S_a^+=\{ \varsigma \in S_a | G(\varsigma) \succ 0 \}$, then $\bar{\varsigma}$ is a global maximizer of $P^d(\varsigma )$ over $S_a^+$, $\bar{x}$ is a global minimizer of $P(x)$ on $\mathbb{R}^n$, and
\begin{eqnarray}
P(\bar{x})=\min_{x\in \mathbb{R}^n} P(x)=\Xi (\bar{x}, \bar{\varsigma}) =\max_{\varsigma \in S_a^+} P^d(\varsigma ) =P^d (\bar{\varsigma}).
\end{eqnarray}
\end{theorem}
It is easy to prove that the canonical dual function $P^d(\varsigma )$ is concave on the convex dual feasible space $S_a^+$. Therefore, Theorem 1 shows that the nonconvex primal problem $(\mathcal{P} )$ is equivalent to a concave maximization problem $(\mathcal{P}^d )$ over a convex space $S_a^+$, which can be solved easily by well-developed methods. Over $S_a^-=\{ \varsigma \in S_a | G(\varsigma) \prec 0 \}$ we have the following theorem:
\begin{theorem} 
(Gao et al., 2012) Suppose that $\bar{\varsigma}$ is a critical point of $(\mathcal{P}^d)$ and the vector $\bar{x}$ is defined by $\bar{x} =G^+(\bar{\varsigma} )F(\bar{\varsigma} )$. If $\bar{\varsigma} \in S_a^-$, then on a neighborhood $\mathcal{X}_o \times \mathcal{S}_o \subset \mathcal{X} \times S_a^-$ of $(\bar{x}, \bar{\varsigma } )$,
we have either
\begin{eqnarray}
P(\bar{x} ) = \min_{x \in \mathcal{X}_o} P(x) =\Xi (\bar{x}, \bar{\varsigma}) =\min_{\varsigma \in \mathcal{S}_o}   P^d(\varsigma ) = P^d(\bar{\varsigma}), \label{equ-dualmin}
\end{eqnarray}
or
\begin{eqnarray}
P(\bar{x} ) = \max_{x \in \mathcal{X}_o} P(x) =\Xi (\bar{x}, \bar{\varsigma}) =\max_{\varsigma \in \mathcal{S}_o}   P^d(\varsigma ) = P^d(\bar{\varsigma}). \label{equ-dualmax}
\end{eqnarray}
\end{theorem}
By the fact that the canonical dual function is a d.c. function (difference of convex functions) on $S_a^-$, the double-min duality (\ref{equ-dualmin}) can be used for finding the biggest local minimizer of $(\mathcal{P} )$ and $(\mathcal{P}^d )$, while the double-max duality (\ref{equ-dualmax}) can be used for finding the biggest local maximizer of $(\mathcal{P} )$ and $(\mathcal{P}^d )$. In physics and material sciences, this pair of biggest local extremal points play important roles in phase transitions.\\

\begin{figure}[h!]
\centerline{
\includegraphics[width=4.2in]{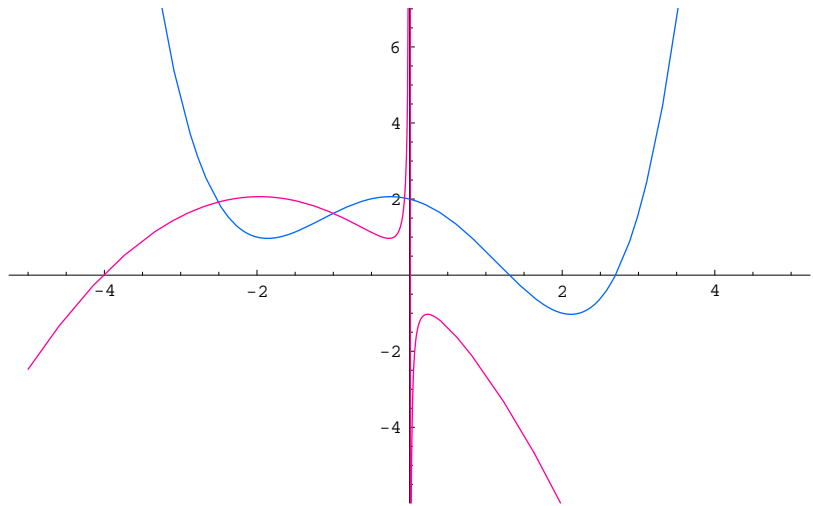}
}
\caption{The prime and dual double-well potential functions (Prime: blue, Dual: red).}
\label{double_well}
\end{figure}

To illuminate that the CDT works, we minimize the well-known Double Well Potential (DWP) function (Gao, 2000) (blue colored in Fig. \ref{double_well}):
\begin{equation}
P(x)=\frac{1}{2} (\frac{1}{2} x^2- 2)^2 -\frac{1}{2} x. \label{double_well_potential_prime}
\end{equation}
\noindent We can easily get $\Xi (x,\varsigma ) =(\frac{1}{2} x^2-2)\varsigma -\frac{1}{2} \varsigma^2 -\frac{1}{2} x$,
\begin{equation}
P^d (\varsigma )=-\frac{1}{8\varsigma }  -\frac{1}{2} \varsigma^2 -2\varsigma \label{double_well_potential_dual}
\end{equation}
(red colored in Fig. \ref{double_well}) and $S_a^+=\{ \varsigma \in \mathbb{R}^1 | \varsigma >0\}$. Let $\Xi (x,\varsigma )'=0$, we get three critical points of $\Xi (x,\varsigma )$: $(\bar{x}^1, \bar{\varsigma}^1)=(2.11491, 0.236417), (\bar{x}^2, \bar{\varsigma}^2)=(-1.86081, -0.268701), (\bar{x}^3, \bar{\varsigma}^3)=(-0.254102,-1.96772)$. By Theorem 1, we know $\bar{x}^1=2.11491$ is the global minimizer of (\ref{double_well_potential_prime}), $\bar{\varsigma}^1=0.236417$ is the global maximizer of (\ref{double_well_potential_dual}) over $S_a^+$, and $P(\bar{\varsigma}^1)=\Xi (\bar{x}^1, \bar{\varsigma}^1)=P^d(\bar{\varsigma}^1)=-1.02951$. By Theorem 2, we know that the local minimizers: $\bar{x}^2=-1.86081, \bar{\varsigma}^2=-0.268701$ (over $S_a^-$), $P(\bar{\varsigma}^2)=\Xi (\bar{x}^2, \bar{\varsigma}^2)=P^d(\bar{\varsigma}^2)=0.9665031$ and the local maximizers: $\bar{x}^3=-0.254102,\bar{\varsigma}^3 =-1.96772$ (over $S_a^-$), $P(\bar{\varsigma}^3)=\Xi (\bar{x}^3, \bar{\varsigma}^3)=P^d(\bar{\varsigma}^3)=2.063$.\\

Thus, by Fig. \ref{double_well} illuminating the application of CDT to the DWP problem, we may see that the canonical dual approach works. The powerful of canonical dual approach is preliminarily shown in Tables 1-3 of arXiv:1105.2270v3: {\tiny 128.84.158.119/PS\_cache/arxiv/pdf/1105/1105.2270v3.pdf} In the next section, we will apply this successful canonical dual approach to the molecular model building and solving problem of prion AGAAAAGA amyloid fibrils.

\section{Prion AGAAAAGA Amyloid Fibril Molecular Model Building and Solving}
Many experimental studies such as (Brown, 2000, Brown, 2001, Brown et al., 1994, Holscher et al., 1998, Jobling et al., 2001, Jobling et al., 1999, Kuwata et al., 2003, Norstrom et al., 2005, Wegner et al., 2002) have shown two points: (1) the hydrophobic region (113-120) AGAAAAGA of prion proteins is critical in the conversion from a soluble PrP$^C$ into an insoluble PrP$^{Sc}$ fibrillar form; and (2) normal AGAAAAGA is an inhibitor of prion diseases. Various computational approaches were used to address the problems related to ``amyloid fibril" (Carter et al., 1998, Chou, 2004b, Chou, 2004c, Chou et al., 2002, Wang et al., 2008, Wei et al., 2005, Zhang, 2011, Zhang et al., 2011, Zhang, 2009). By introducing novel mathematical canonical dual formulations and computational approaches, in this paper we may construct atomic-resolution molecular structures for prion (113–-120) AGAAAAGA amyloid fibrils.\\ 

The atomic structures of all amyloid fibrils revealed steric zippers, with strong van der Waals interactions between $\beta$-sheets and hydrogen bonds to maintain the $\beta$-strands (Sawaya et al., 2007). About $\beta$-sheets and $\beta$-barrels, there are various interactions and motions, such as the interactions between $\beta$-strands (Chou et al., 1982b, Chou et al., 1982a, Chou et al., 1983a, Chou et al., 1983b), interaction between two $\beta$-sheets (Chou et al., 1986), as well as the low-frequency accordion-like motion in a $\beta$-sheet and breathing motion in a $\beta$-barrel (Chou, 1985) and their biological functions (Chou, 1988). The ``amyloid fibril" problem can be looked as a molecular distance geometry problem (MDGP) (Grosso et al., 2009), which arises in the interpretation of NMR data and in the determination of protein structure [as an example to understand MDGP, the problem of locating sensors in telecommunication networks is a DGP. In such a case, the positions of some sensors are known (which are called anchors) and some of the distances between sensors (which can be anchors or not) are known: the DGP is to locate the positions of all the sensors. Here we look sensors as atoms and their telecommunication network as a molecule]. The three dimensional structure of a molecule with $n$ atoms can be described by specifying the 3-Dimensional coordinate positions $x_1, x_2, \dots, x_n \in \mathbb{R}^3$ of all its atoms. Given bond lengths $d_{ij}$ between a subset $S$ of the atom pairs, the determination of the molecular structure is
\begin{eqnarray}
(\mathcal{P}_0 ) \quad to \quad find \quad &x_1,x_2,\dots ,x_n  \quad s.t. \quad ||x_i-x_j||=d_{ij}, (i,j)\in S,  \label{orginal_problem}
\end{eqnarray}
where $||\cdot ||$ denotes a norm in a real vector space and it is calculated as the Euclidean distance 2-norm in this paper. (\ref{orginal_problem}) can be reformulated as a mathematical global optimization problem (GOP)
\begin{eqnarray}
(\mathcal{P} ) \quad &\min P(X)=\sum_{(i,j)\in S} w_{ij} (||x_i-x_j||^2 -d_{ij}^2 )^2  \label{prime_problem}
\end{eqnarray}
in the terms of finding the global minimum of the function $P(X)$, where $w_{ij}, (i,j)\in S$ are positive weights, $X = (x_1, x_2, \dots, x_n)^T \in \mathbb{R}^{n\times 3}$ (More et al., 1997) and usually $S$ has many fewer than $n^2/2$ elements due to the error in the theoretical or experimental data (Zou et al., 1997, Grosso et al., 2009). There may even not exist any solution $x_1, x_2, \dots, x_n$ to satisfy the distance constraints in (\ref{orginal_problem}), for example when data for atoms $i, j, k \in S$ violate the triangle inequality; in this case, we may add a perturbation term $-\epsilon^TX$ to $P(X)$:
\begin{eqnarray}
(\mathcal{P}_{\epsilon} ) \quad &\min P_{\epsilon}(X)=\sum_{(i,j)\in S} w_{ij} (||x_i-x_j||^2 -d_{ij}^2 )^2 -\epsilon^TX, \label{prime_approx_problem}
\end{eqnarray}
where $\epsilon \geq 0$. In some cases, instead exact values $d_{ij}, (i,j)\in S$ can be found, we can only specify lower and upper bounds on the distances: $l_{ij} \leq ||x_i - x_j || \leq u_{ij} , (i, j) \in S$; in such cases we may penalize all the unsatisfied constraints into the objective function of ($\mathcal{P}_{\epsilon}$) by adding
$\sum_{(i,j)\in S} \left( \max \left\{ l^2_{i,j}- ||x_i - x_j ||^2,  0\right\} \right)^2 
                 + \left( \max \left\{ ||x_i - x_j ||^2 - u^2_{i,j}, 0\right\} \right)^2$ into $P_{\epsilon}(X)$ (Zou et al., 1997, Grosso et al., 2009), where we may let $d_{ij}$ be the interatomic distance (less than 6 angstroms) for the pair in successive residues of a protein and set $l_{ij}=(1-0.05)d_{ij}$ and $u_{ij}=(1+0.05) d_{ij}$ (Grosso et al., 2009). In this paper we will use the canonical duality approach introduced in Section 2 (Gao et al., 2010,
Gao et al., 2012, Gao, 2000) to solve (\ref{orginal_problem})-(\ref{prime_approx_problem}). Because the canonical dual is a perfect dual with zero duality gap between prime and dual problems, we can get the accurate global optimal solutions of problems (\ref{orginal_problem})-(\ref{prime_approx_problem}). Thus by canoncial dual approach we may successfully construct the molecular structure of prion AGAAAAGA amyloid fibrils as follows.\\

\begin{figure}[h!]
\centerline{
\includegraphics[scale=0.6]{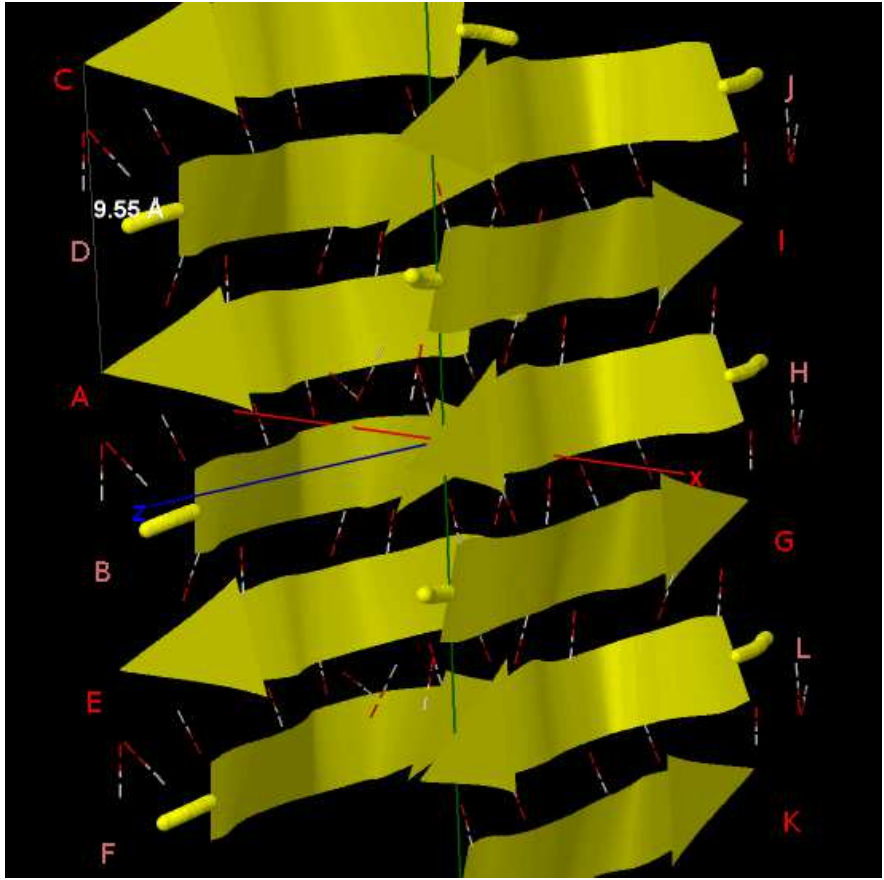}
}
\caption{Protein fibril structure of human M129 prion GYMLGS (127--132).}
\label{fig_3NHC}
\end{figure}

\begin{figure}[h!]
\centerline{
\includegraphics[scale=0.6]{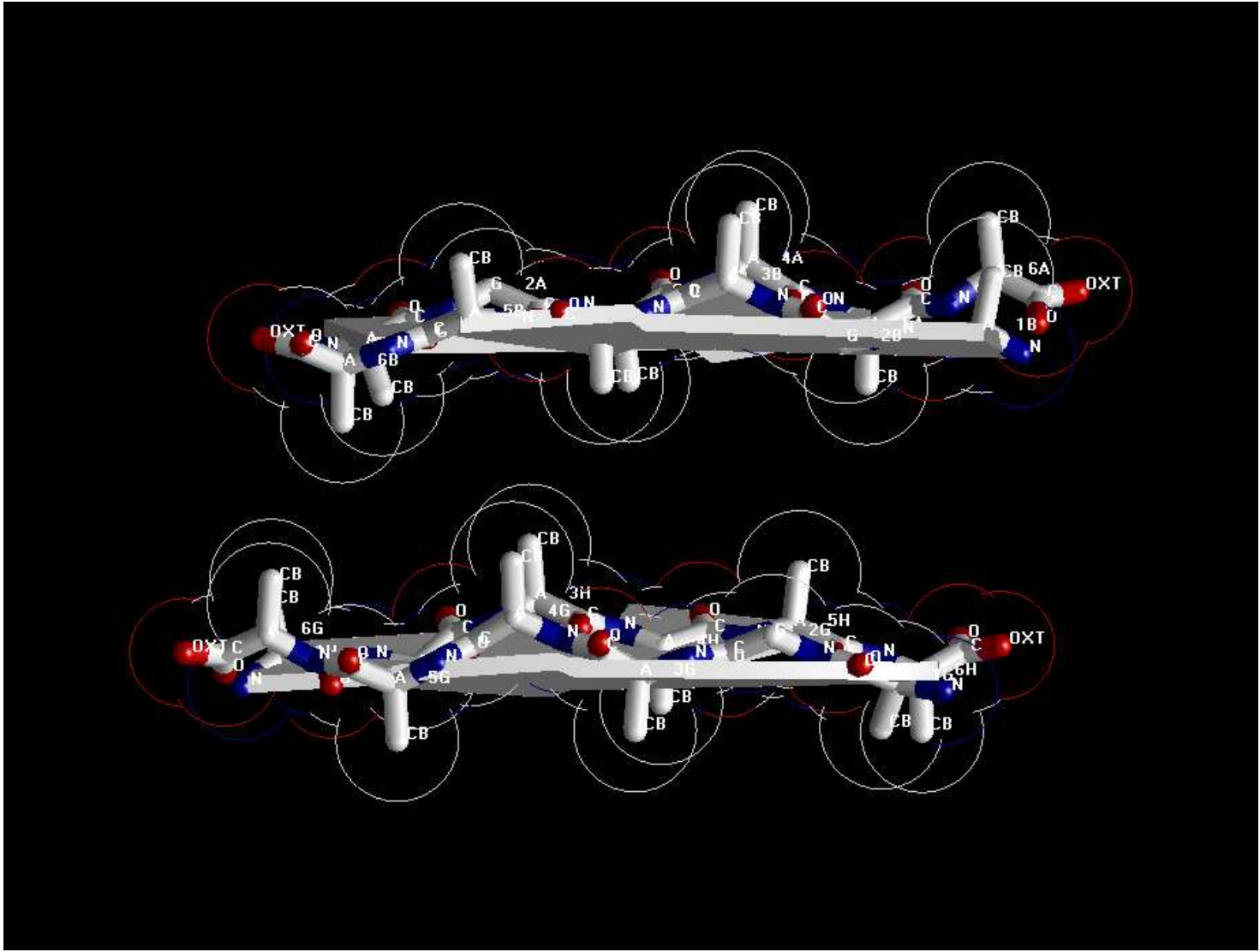}
}
\caption{Far vdw contacts of AG chains and BH chains of Model 1.}
\label{fig_bad_VDW_1}
\end{figure}

\begin{figure}[h!]
\centerline{
\includegraphics[scale=0.6]{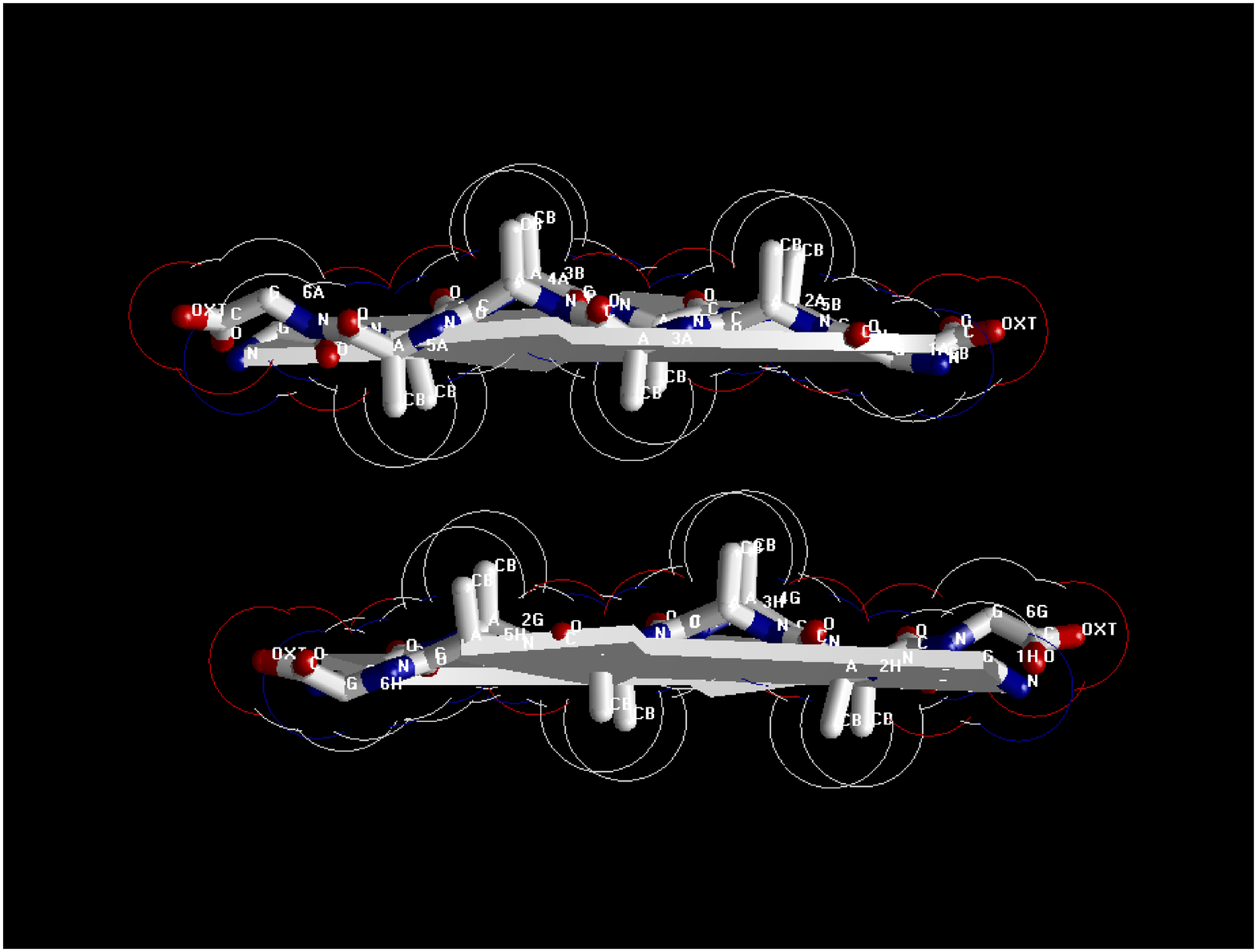}
}
\caption{Far vdw contacts of AG chains and BH chains of Model 2.}
\label{fig_bad_VDW_2}
\end{figure}

\begin{figure}[h!]
\centerline{
\includegraphics[scale=0.6]{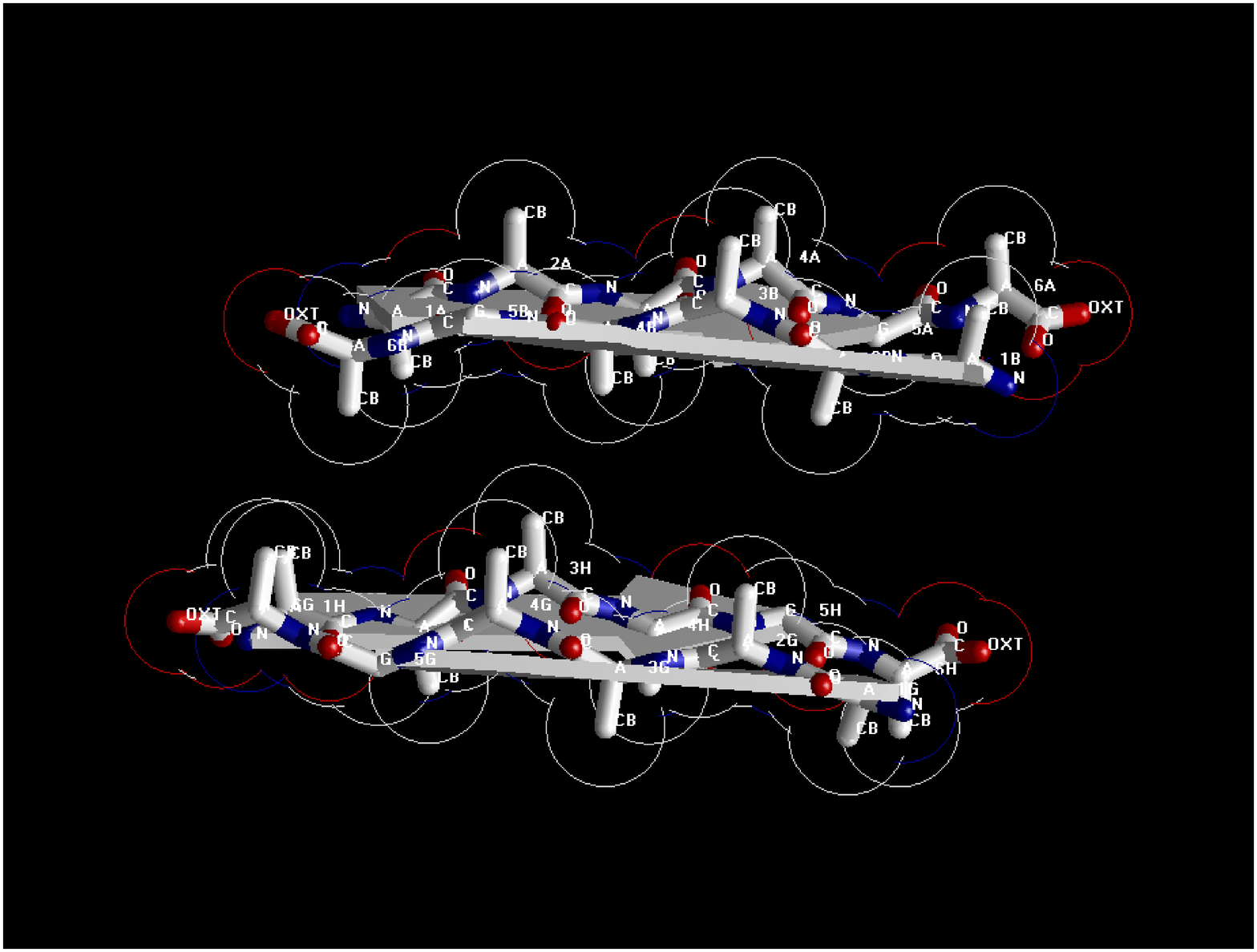}
}
\caption{Far vdw contacts of AG chains and BH chains of Model 3.}
\label{fig_bad_VDW_3}
\end{figure}

\begin{figure}[h!]
\centerline{
\includegraphics[scale=0.6]{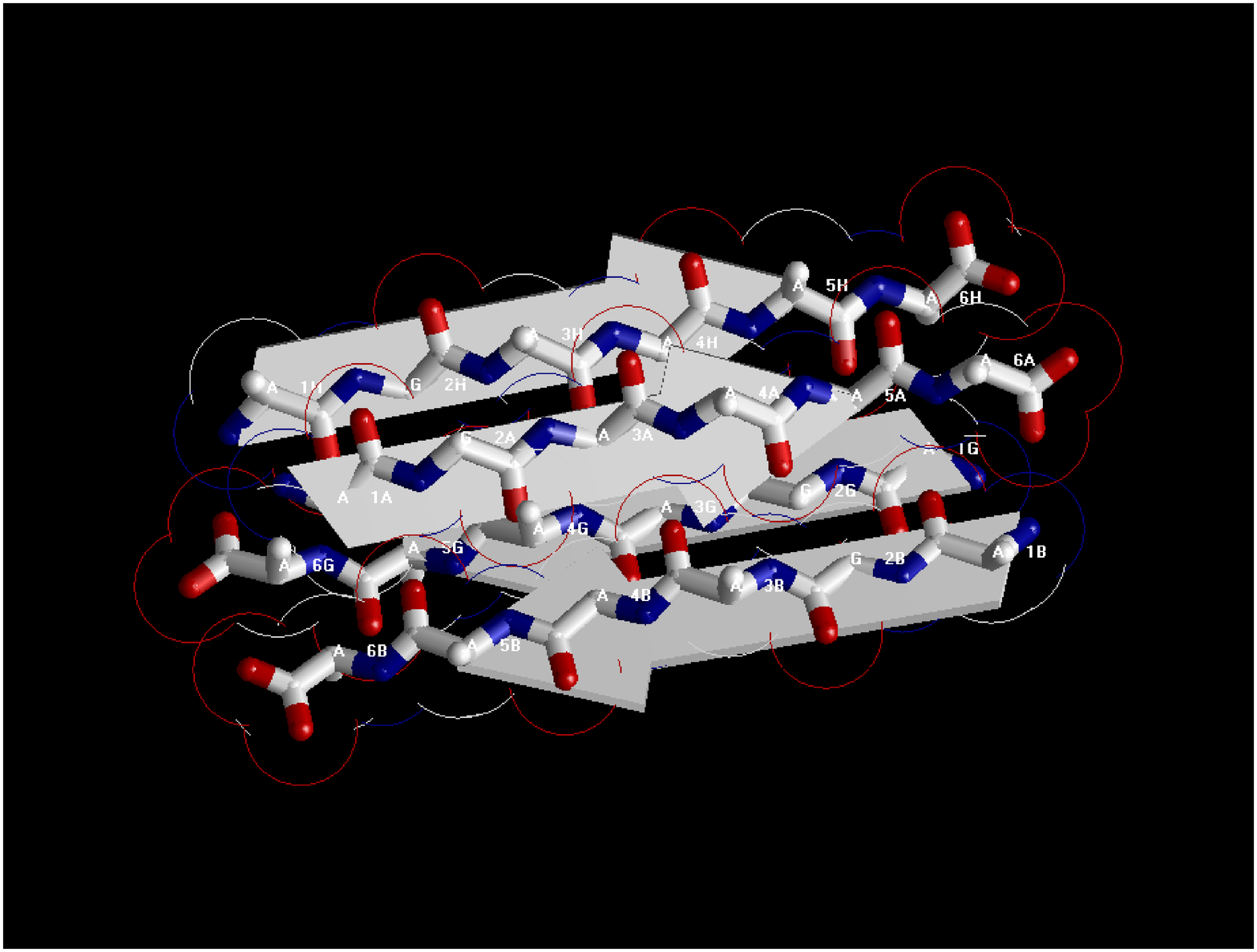}
}
\caption{Close vdw contacts of AG chains and BH chains of Model 1.}
\label{fig_good_VDW_1}
\end{figure}

\begin{figure}[h!]
\centerline{
\includegraphics[scale=0.6]{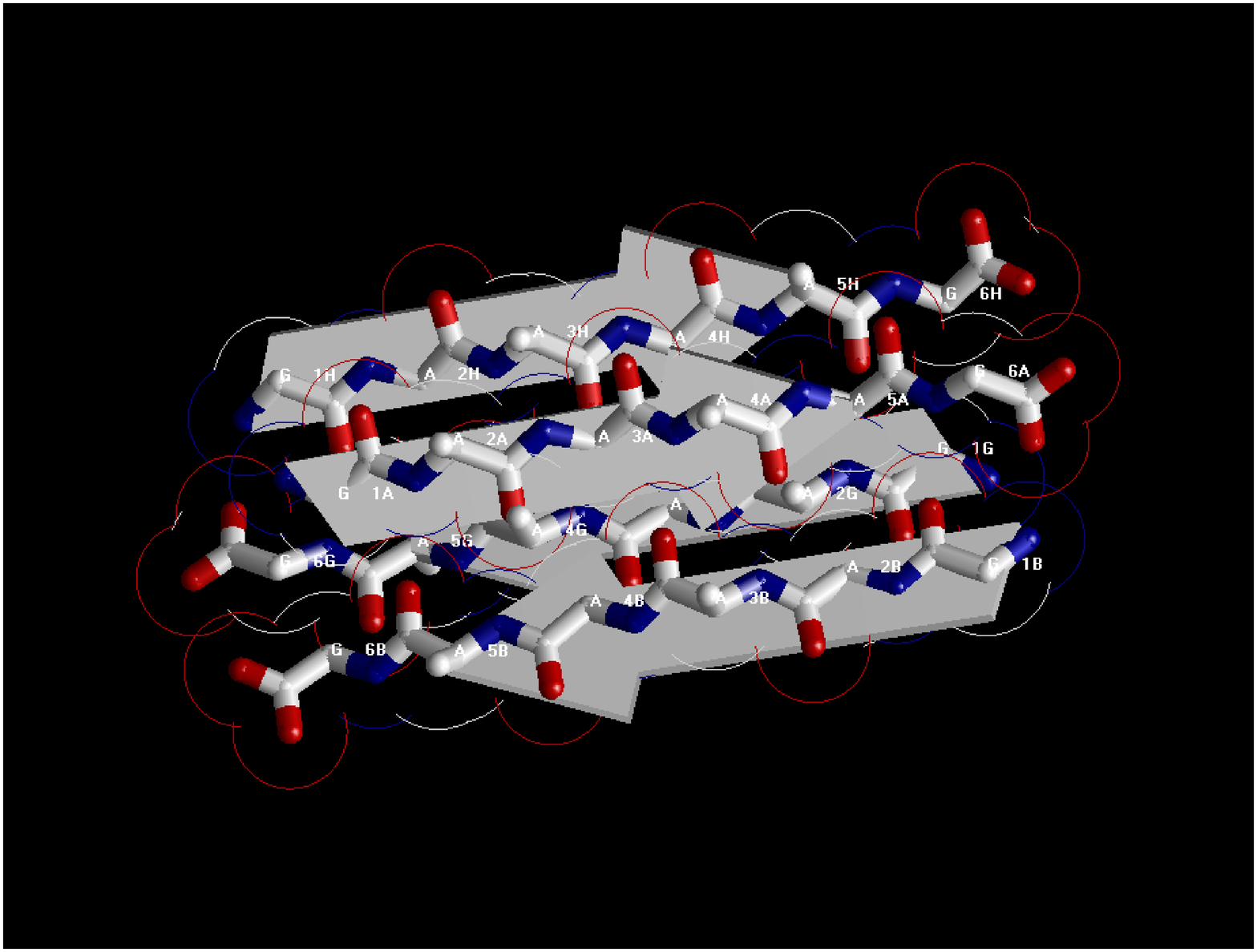}
}
\caption{Close vdw contacts of AG chains and BH chains of Model 2.}
\label{fig_good_VDW_2}
\end{figure}

\begin{figure}[h!]
\centerline{
\includegraphics[scale=0.6]{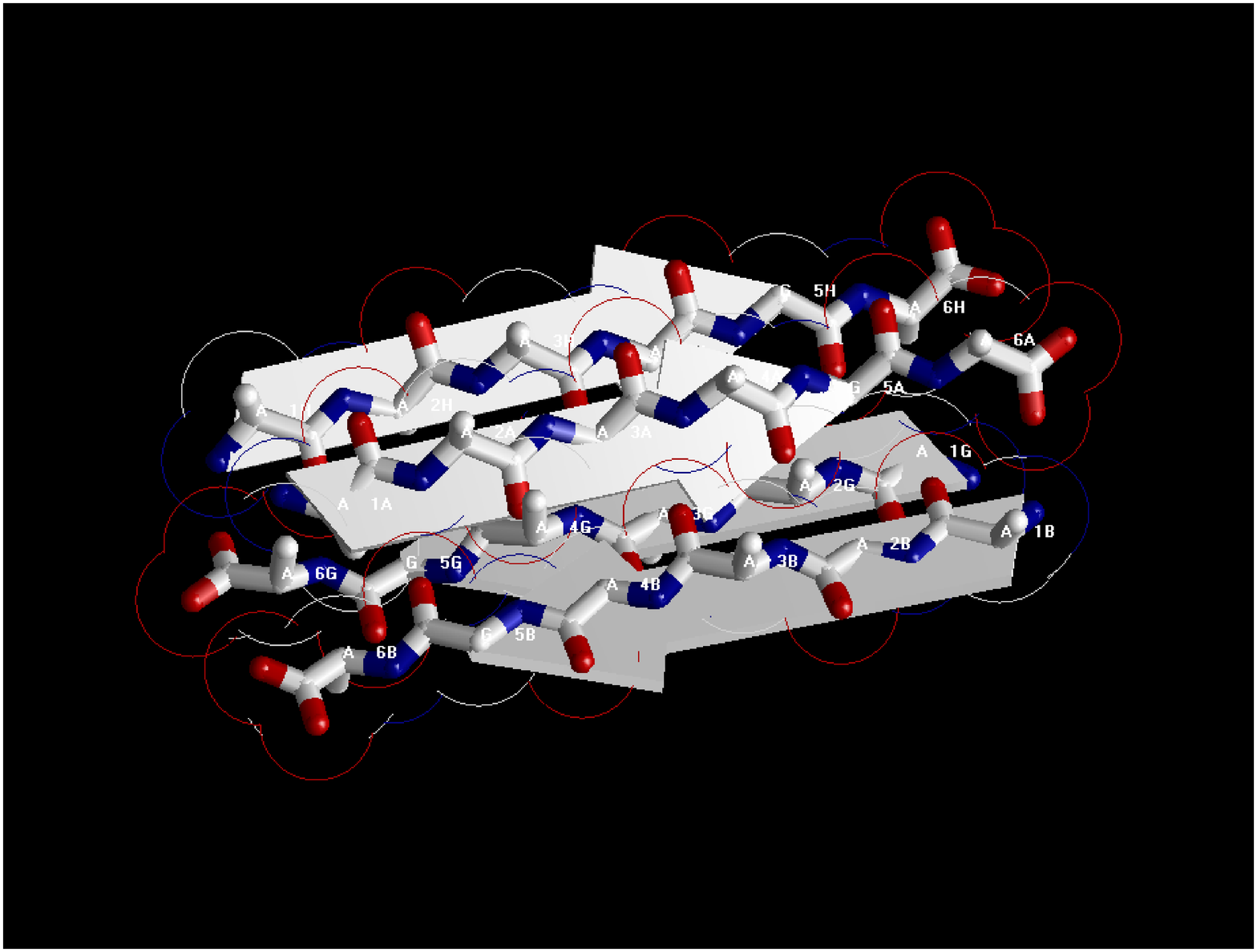}
}
\caption{Close vdw contacts of AG chains and BH chains of Model 3.}
\label{fig_good_VDW_3}
\end{figure}

\begin{figure}[h!]
\centerline{
\includegraphics[scale=0.6]{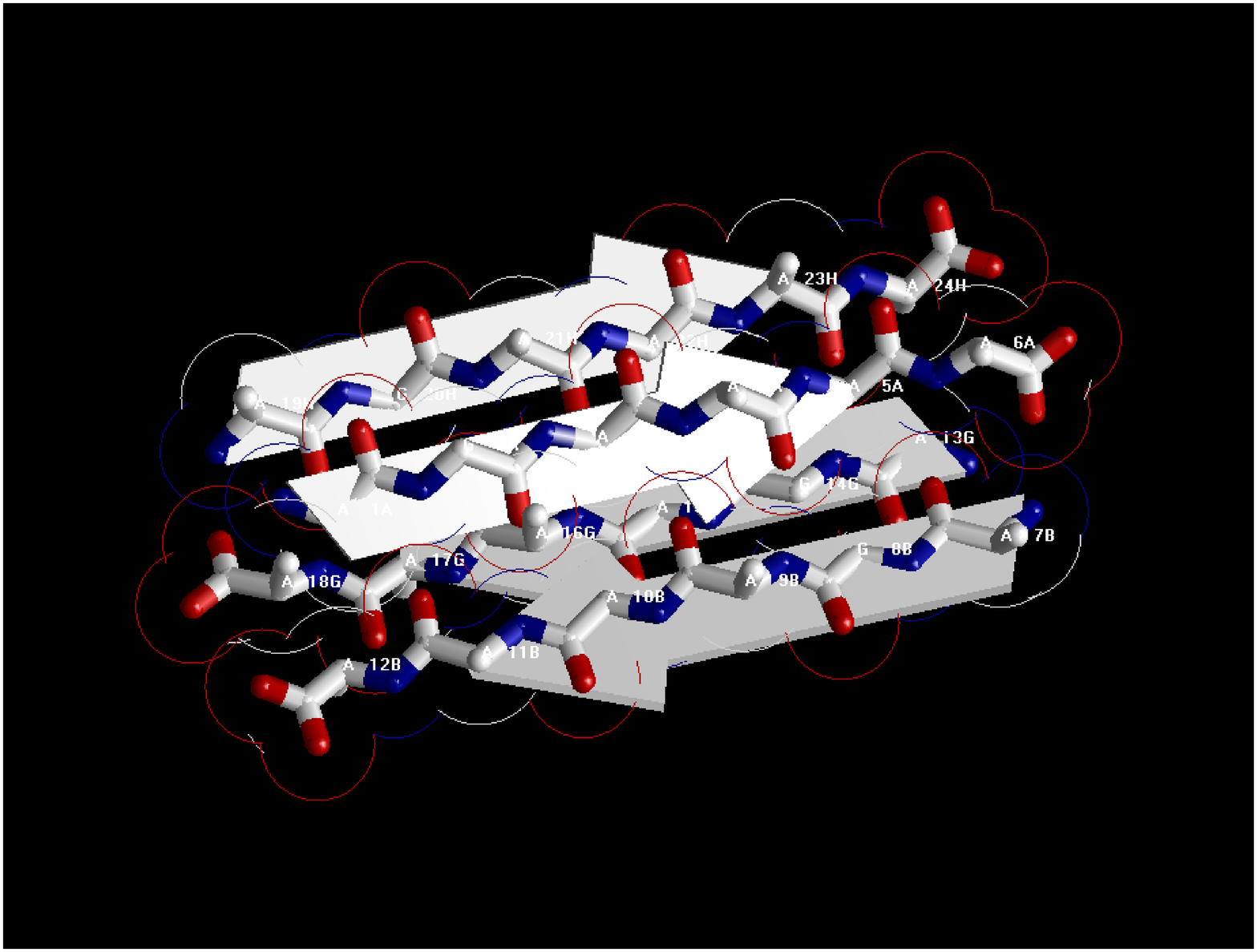}
}
\caption{Optimal structure of prion AGAAAAGA amyloid fibril Model 1.}
\label{fig_good_AMBER_1}
\end{figure}

\begin{figure}[h!]
\centerline{
\includegraphics[scale=0.6]{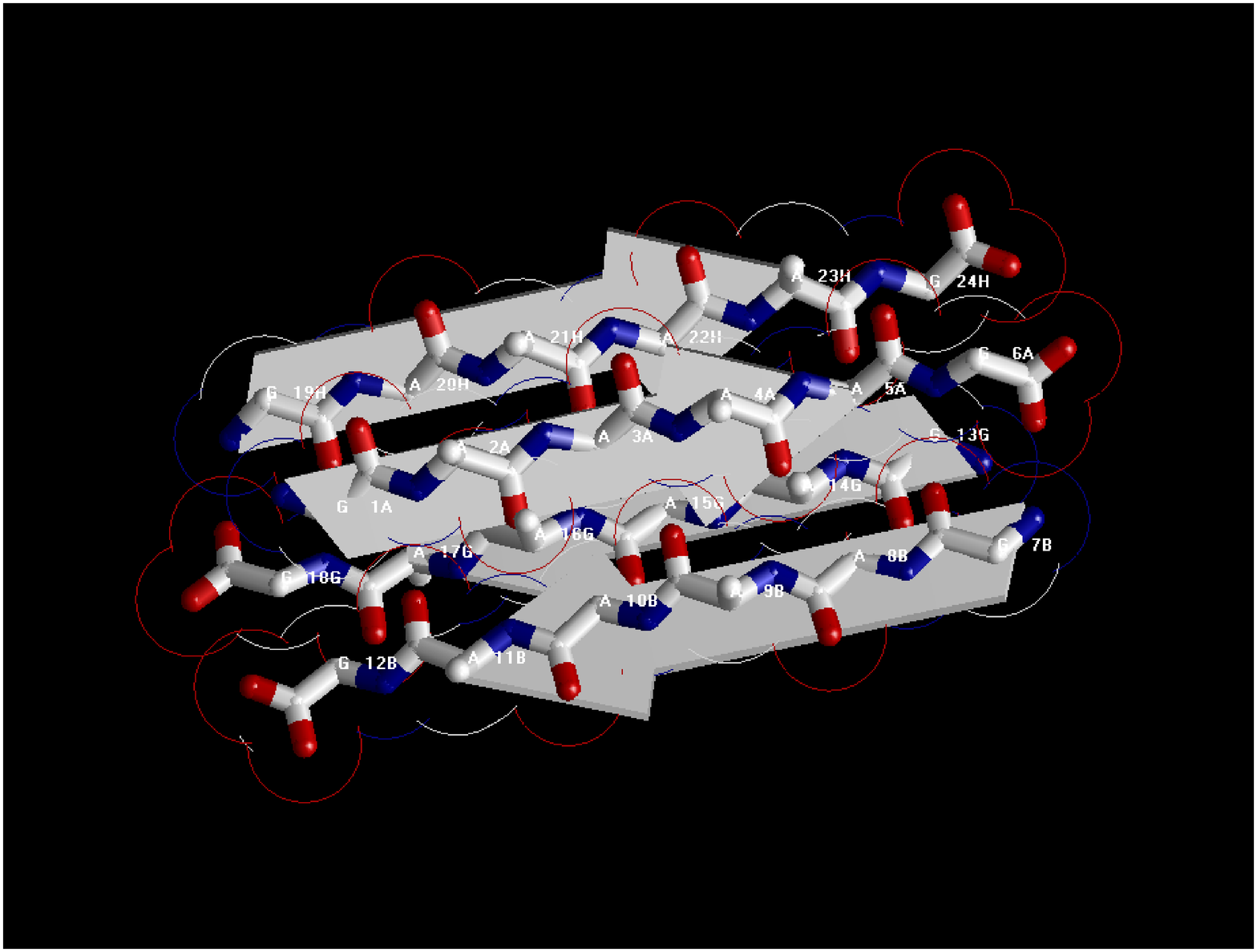}
}
\caption{Optimal structure of prion AGAAAAGA amyloid fibril Model 2.}
\label{fig_good_AMBER_2}
\end{figure}

\begin{figure}[h!]
\centerline{
\includegraphics[scale=0.6]{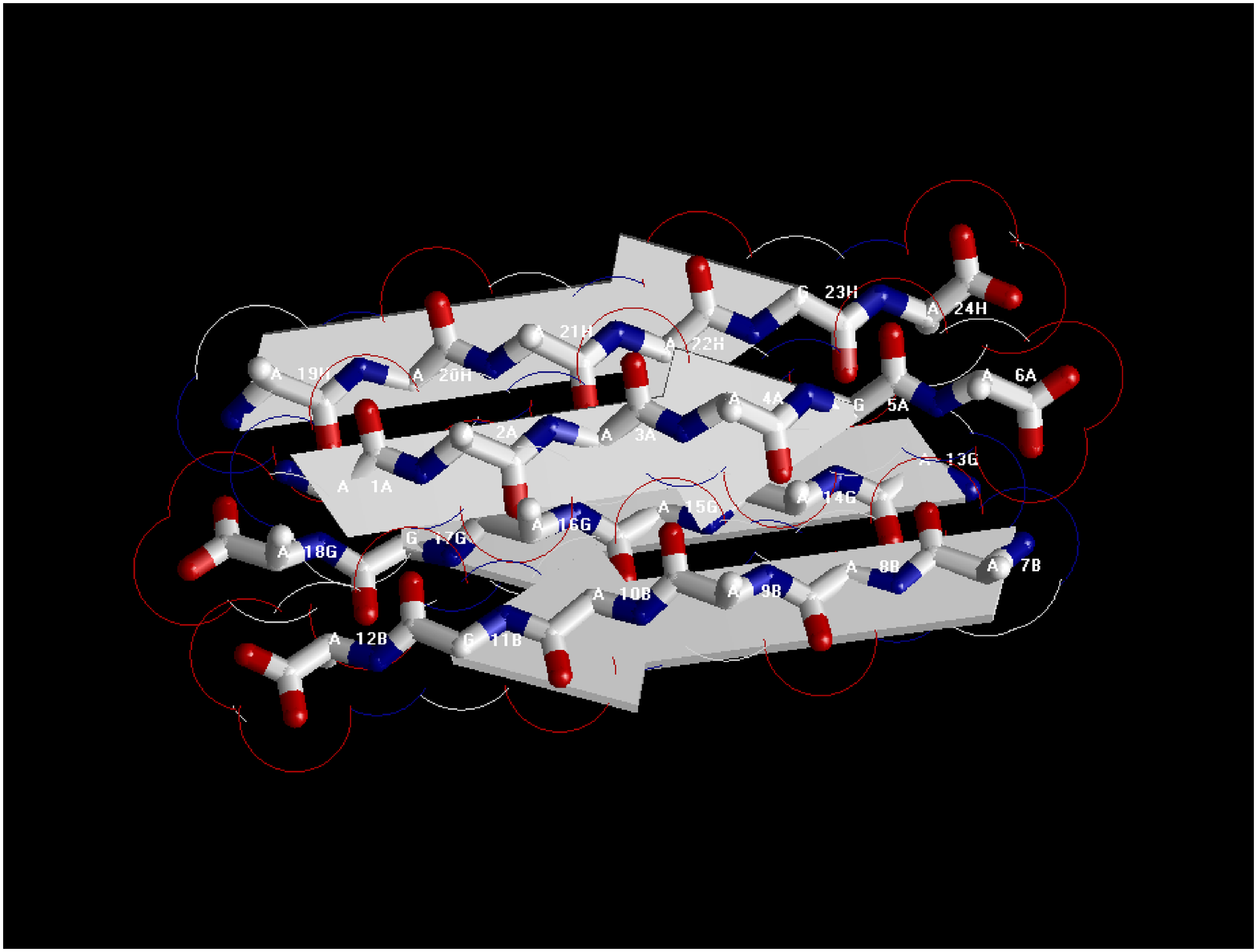}
}
\caption{Optimal structure of prion AGAAAAGA amyloid fibril Model 3.}
\label{fig_good_AMBER_3}
\end{figure}

If we look at the prion AGAAAAGA molecular modeling problem as a MDGP with two anchors and two sensors, we can easily construct the prion AGAAAAGA amyloid fibril models. In fact we may let the coordinates of these two anchors being variable. But, these two anchors belong to one body of Chains A and B, and the two sensors belong to another body of Chains G and H. This is a simple two-body problem model of theoretical physics, i.e. Einstein's absolute relative theory. Hence, we may look the coordinates of two anchors being fixed. The constructions will be based on the most recently released experimental molecular structures of human M129 prion peptide 127–-132 (PDB entry 3NHC released into Protein Data Bank (http://www.rcsb.org) on 04-AUG-2010) (in brief, this paper will use the PrP structured region 127-132 to do homology modelling for the PrP unstructured region 113-120). The atomic-resolution structure of this peptide is a steric zipper, with strong van der Waals (vdw) interactions between $\beta$-sheets and hydrogen bonds to maintain the $\beta$-strands (Fig. \ref{fig_3NHC}, where the dashed lines denote the hydrogen bonds).
\noindent In Fig. \ref{fig_3NHC} we see that G (H) chains (i.e. $\beta$-sheet 2) of 3NHC.pdb can be obtained from A (B) chains (i.e. $\beta$-sheet 1) by
\begin{equation}
G (H) = \left( \begin{array}{ccc}
 1  &0  &0  \\
 0  &-1 &0  \\
 0  &0  &-1 \end{array} \right) A (B) + 
\left( \begin{array}{c}
 9.07500\\
 4.77650\\
 0.00000 
\end{array} \right), \label{gh}
\end{equation}
and other chains can be got by
\begin{equation}
I (J) = G (H) +\left( \begin{array}{c} 
0\\
9.5530\\ 
0\end{array} \right), 
K (L) = G (H) + \left( \begin{array}{c} 
0\\
-9.5530\\ 
0\end{array} \right), \label{ijkl}
\end{equation}
\begin{equation}
C (D)= A (B)+ \left( \begin{array}{c}
0\\
9.5530\\ 
0\end{array} \right),
E (F) = A (B) +\left( \begin{array}{c}
0\\
-9.5530\\ 
0\end{array} \right). \label{cdef}
\end{equation}
Basing on the template 3NHC.pdb from the Protein Data Bank, three prion AGAAAAGA palindrome amyloid fibril models –- an AGAAAA model (Model 1), a GAAAAG model (Model 2), and an AAAAGA model (Model 3) –- will be successfully constructed in this paper. AB chains of Models 1-3 were respectively got from AB chains of 3NHC.pdb using the mutate module of the free package Swiss-PdbViewer (SPDBV Version 4.01) ({\small http://spdbv.vital-it.ch}). It is pleasant to see that almost all the hydrogen bonds are still kept after the mutations, where for the donor O (oxygen) atom and the acceptor H (hydrogen) atom if the distance cutoff is less than 3.00 angstroms and the angle cutoff is less than 120.00 degrees then a hydrogen bond is kept; thus we just need to consider the vdw contacts only. Making mutations for GH chains of 3NHC.pdb, we can get the GH chains of Models 1-3. However, the vdw contacts between A chain and G chain, between B chain and H chain are too far at this moment (Fig.s \ref{fig_bad_VDW_1}-\ref{fig_bad_VDW_3}) because the shortest distance of atoms between Chain A and Chain G, and between Chain B and Chain H, is still very larger than the double size of the vdw radius of CB carbon atom.
\noindent Seeing Fig.s \ref{fig_bad_VDW_1}-\ref{fig_bad_VDW_3}, we may know that for Models 1-3 at least two vdw interactions between A.ALA3.CB-G.ALA4.CB, B.ALA4.CB-H.ALA3.CB should be maintained. Fixing the coordinates of A.ALA3.CB and B.ALA4.CB (two anchors) ((6.014,5.917,0.065), (5.658,1.630,-0.797)), letting $d$ equal to the twice of the vdw radius of Carbon atom (i.e. $d =3.4$ angstroms), and letting the coordinates of G.ALA4.CB and H.ALA3.CB (two sensors) be variables, we may get a simple MDGP with 6 variables and its dual with 2 variables:
\begin{eqnarray*}
P(x_1,x_2)&=&\frac{1}{2} \left\{ (x_{11} -6.014)^2+(x_{12}-5.917)^2 +(x_{13}- 0.065)^2 -3.4^2 \right\}^2\\ 
          &+&\frac{1}{2} \left\{ (x_{21} -5.658)^2+(x_{22}-1.630)^2 +(x_{23}+ 0.797)^2 -3.4^2 \right\}^2,\\
P^d(\varsigma_1, \varsigma_2) &=&-11.56\varsigma_1 -\frac{1}{2} \varsigma_1^2 
                                 -11.56\varsigma_2 -\frac{1}{2} \varsigma_2^2.
\end{eqnarray*}
We can get a local maximal solution (-11.56,-11.56) for $P^d(\varsigma_1, \varsigma_2)$ and its corresponding local maximal solution to $P(x_1,x_2)$. But we need the global maximal solution of $P^d(\varsigma_1, \varsigma_2)$. Thus, by introducing perturbation parameters $\epsilon =0.05$, we have to seek the global optimal solutions from the perturbed problems of $P(x_1,x_2)$ and $P^d(\varsigma_1, \varsigma_2)$:
\begin{eqnarray*}
P_{\epsilon}(x_1,x_2)&=& \frac{1}{2} \left\{ (x_{11} -6.014)^2+(x_{12}-5.917)^2 +(x_{13}- 0.065)^2 -3.4^2 \right\}^2\\
          &+&\frac{1}{2} \left\{ (x_{21} -5.658)^2+(x_{22}-1.630)^2 +(x_{23}+ 0.797)^2 -3.4^2 \right\}^2\\
          &-&0.05 x_{11} -0.05x_{12} -0.05x_{13} -0.05x_{21} -0.05x_{22} -0.05x_{23},\\
 P_{\epsilon}^d(\varsigma_1, \varsigma_2) &=&59.6233\varsigma_1 -0.5\varsigma_1^2 + 23.7451\varsigma_2 -0.5\varsigma_2^2\\
&-&\frac{1}{2} \left( 
\frac{(0.05+12.028\varsigma_1)^2}{2\varsigma_1} +
\frac{(0.05+11.834\varsigma_1)^2}{2\varsigma_1} +
\frac{(0.05+ 0.130\varsigma_1)^2}{2\varsigma_1}
 \right)\\ 
&-&\frac{1}{2} \left(
\frac{(0.05+11.316\varsigma_2)^2}{2\varsigma_2} +
\frac{(0.05+3.2600\varsigma_2)^2}{2\varsigma_2} +
\frac{(0.05- 1.594\varsigma_2)^2}{2\varsigma_2}
\right).                
\end{eqnarray*}
We can easily get the global maximal solution $(0.0127287, 0.0127287) \in \{ \varsigma \in \mathbb{R}^2 | \varsigma_i >0, i=1,2\}$ for $P_{\epsilon}^d(\varsigma_1, \varsigma_2)$. Then, we get its corresponding solution for $P_{\epsilon}(x_1,x_2)$:\\
\centerline{$\bar{x}=(7.97807, 7.88107, 2.02907, 7.62207, 3.59407, 1.16707).$}
By Theorem 1 we know that $\bar{x}$ is a global minimal solution of $P_{\epsilon}(x_1,x_2)$. We set $\bar{x}$ as the coordinates of G.ALA4.CB and H.ALA3.CB and taking the average value we get
\begin{equation}
G (H) = \left( \begin{array}{ccc}
 1  &0  &0  \\
 0  &-1 &0  \\
 0  &0  &-1 \end{array} \right) A (B) + 
\left( \begin{array}{c}
 1.96407\\
 9.51107\\
 1.23207 
\end{array} \right). \label{gh_last}
\end{equation}
By (\ref{gh_last}) we can get very close vdw contacts between A chain and G chain, between B chain and H chain (Fig.s \ref{fig_good_VDW_1}-\ref{fig_good_VDW_3}). 
\noindent Thus, we successfully constructed Models 1-3, and through further refinements by the Amber 11 package (Case et al., 2010) we at last get the optimal Models (Fig.s \ref{fig_good_AMBER_1}-\ref{fig_good_AMBER_3}). We find the RMSD (root mean square deviation) between Fig.s \ref{fig_good_VDW_1}-\ref{fig_good_VDW_3} and Fig.s \ref{fig_good_AMBER_1}-\ref{fig_good_AMBER_3} is zero angstroms; this implies that the Amber 11 refinements are not necessary and the CDT is good enough to get the optimal Models 1-3 as illuminated in Fig.s \ref{fig_good_VDW_1}-\ref{fig_good_VDW_3}. The other CDIJ and EFKL chains can be got by parallelizing ABGH chains in the use of mathematical formulas (\ref{ijkl})-(\ref{cdef}).

As the end of this Section, we give some remarks on the Models 1-3. (1) The canonical dual approach exactly makes the closest CB atoms between Chain A and Chain G, and between Chain B and Chain H, just being equal to the double size of the vdw radius of CB carbon atom (Fig.s 6-8) and this is the perfect structure of the Models 1-3. Fig.s 9-11 were got by the further refinements through the SDCG optimization methods of Amber 11 package. The zero RMSD value between Fig.s 6-8 and Fig.s 9-11 implies to us that the canonical dual approach of this paper works well. (2) The SDCG optimization methods of Amber 11 package automatically considered the bond angles and dihedral angles, and during the canonical dual molecular model building and optimization procedure, the perfect bond angles and dihedral angles automatically produced by the Swiss-PdbViewer package are still being kept. (3) The molecular modeling problem of this paper is in fact a very simple two-body problem of theoretical physics, i.e. Einstein's absolute relative theory. In mathematics, it is a sensor network problem with two anchors and two sensors. 

\section{Conclusion}
This paper presents an important method and provides useful information for treatments of prion diseases. X-ray crystallography is a powerful tool to determine the protein 3D structure. However, it is time-consuming and expensive, and not all proteins can be successfully crystallized, particularly for membrane proteins. Although NMR spectroscopy is indeed a very powerful tool in determining the 3D structures of membrane proteins, it is also time-consuming and costly. Due to the noncrystalline and insoluble nature of the neurodegenerative amyloid fibril or plaque, little structural data on the prion AGAAAAGA segment is available. Under these circumstances, the novel canonical dual computational approach introduced in this paper showed its power in the molecular modeling of prion AGAAAAGA amyloid fibrils. This indicated that computational approaches or introducing novel mathematical formulations and physical concepts into molecular biology can significantly stimulate the development of biological and medical science. The optimal atomic-resolution structures of prion AGAAAAGA amyloid fibils presented in this paper are useful for the drive to find treatments for prion diseases in the field of medicinal chemistry.\\

\noindent {\bf Acknowledgments:} This research is supported by a Victorian Life Sciences Computation Initiative (http://www.vlsci.org.au) grant number VR0063 on its Peak Computing Facility at the University of Melbourne, an initiative of the Victorian Government. This research is supported by US Air Force Office of Scientific Research under the grant AFOSR FA9550-10-1-0487. The second author contributed to check his canonical dual theory of Section 2 and gave some suggestions to replace the Rosenbrock Problem of the earlier version of this manuscript. The third author has given numerous supports from the School. The first author is certainly responsible for all faults if exist in the paper. Last, but not the least, the authors appreciate the anonymous referees for their numerous insightful comments to improve this paper, and appreciate the great helps from Ms. Janet Stein, the Journal Manager of Journal of Theoretical Biology.

\section*{References}
Bagirov, A.M., Karasozen, B., Sezer, M., 2008. Discrete gradient method: a derivative free method for nonsmooth optimization. J. Opt. Theor. Appl. 137, 317-–334.\\

\noindent Brown, D.R., 2000. Prion protein peptides: optimal toxicity and peptide blockade of toxicity. Mol. Cell. Neurosci. 15, 66–-78.\\

\noindent Brown, D.R., 2001. Microglia and prion disease. Microsc. Res. Tech. 54, 71-–80.\\

\noindent Brown, D.R., Herms, J., Kretzschmar, H.A., 1994. Mouse cortical cells lacking cellular PrP survive in culture with a neurotoxic PrP fragment. Neuroreport 5, 2057--2060.\\

\noindent Cai, L., Wang, Y., Wang, J.F., Chou, K.C., 2011. Identification of proteins interacting with human SP110 during the process of viral infections. Med. Chem. 7, 121--126.\\

\noindent Call, M. E., Wucherpfennig, K. W., Chou, J. J., 2010. The structural basis for intramembrane assembly of an activating immunoreceptor complex. Nat. Immunol. 11, 1023--1029.\\

\noindent Carter, D.B., Chou, K.C., 1998. A model for structure dependent binding of Congo Red to Alzeheimer beta-amyloid fibrils. Neurobiol. Aging 19, 37--40.\\

\noindent Case, D.A., Darden, T.A., Cheatham, T.E., Simmerling, III C.L., Wang, J., Duke, R.E., Luo, R., Walker, R.C., Zhang, W., Merz, K.M., Roberts, B.P., Wang, B., Hayik, S., Roitberg, A., Seabra, G., Kolossváry, I., Wong, K.F., Paesani, F., Vanicek, J., Liu, J., Wu, X., Brozell, S.R., Steinbrecher, T., Gohlke, H., Cai, Q., Ye, X., Wang, J., Hsieh, M.-J., Cui, G., Roe, D.R., Mathews, D.H., Seetin, M.G., Sagui, C., Babin, V., Luchko, T., Gusarov, S., Kovalenko, A., Kollman, P.A., 2010. AMBER 11, University of California, San Francisco.\\

\noindent Chen, C., Chen, L., Zou, X., Cai, P., 2009. Prediction of protein secondary structure content by using the concept of Chou's pseudo amino acid composition and support vector machine. Protein Peptide Lett. 16, 27--31.\\

\noindent Chou, K. C., 1985. Low-frequency motions in protein molecules: beta-sheet and beta-barrel. Biophys. J. 48, 289--297.\\

\noindent Chou, K. C., 1988. Review: Low-frequency collective motion in biomacromolecules and its biological functions. Biophys. Chem. 30, 3--48.\\

\noindent Chou, K.C., 2004. Insights from modelling the tertiary structure of BACE2. J. Proteome Res. 3, 1069--72.\\

\noindent Chou, K. C., 2004. Modelling extracellular domains of GABA-A receptors: subtypes 1, 2, 3, and 5. Biochem. Biophys. Res. Commun. 316, 636--642.\\

\noindent Chou, K. C., 2004. Molecular therapeutic target for type-2 diabetes. J. Proteome Res. 3, 1284--1288.\\

\noindent Chou, K. C., 2004. Insights from modelling the 3D structure of the extracellular domain of alpha7 nicotinic acetylcholine receptor. Biochem. Biophy. Res. Co. 319, 433--438.\\

\noindent Chou, K.C., 2004. Review: structural bioinformatics and its impact to biomedical science. Curr. Med. Chem. 11, 2105--34.\\

\noindent Chou, K.C., 2005. Modeling the tertiary structure of human cathepsin-E. Biochem. Biophys. Res. Commun. 331, 56--60.\\

\noindent Chou, K. C., 2005. Coupling interaction between thromboxane A2 receptor and alpha-13 subunit of guanine nucleotide-binding protein. J. Proteome Res. 4, 1681--1686.\\

\noindent Chou, K.C., 2011. Some remarks on protein attribute prediction and pseudo amino acid composition (50th Anniversary Year Review). J. Theor. Biol. 273, 236--247.\\

\noindent Chou, K.C., Carlacci, L., 1991. Energetic approach to the folding of alpha/beta barrels. Proteins: Struct., Funct., Genet. 9, 280--295.\\

\noindent Chou, K.C., Carlacci, L., Maggiora, G.M., 1990b. Conformational and geometrical properties of idealized beta-barrels in proteins. J. Mol. Biol. 213, 315--326.\\

\noindent Chou, K.C., Howe, W.J., 2002. Prediction of the tertiary structure of the beta-secretase zymogen. Biochem. Biophys. Res. Commun. 292, 702--8.\\

\noindent Chou, K. C., Nemethy, G., Rumsey, S., Tuttle, R. W., Scheraga, H. A., 1986. Interactions between two beta-sheets: Energetics of beta/beta packing in proteins. J. Mol. Biol. 188, 641--649.\\

\noindent Chou, K.C., Nemethy, G., Scheraga, H.A., 1983. Effects of amino acid composition on the twist and the relative stability of parallel and antiparallel beta-sheets. Biochemistry 22, 6213--6221.\\

\noindent Chou, K. C., Nemethy, G., Scheraga, H. A., 1983. Role of interchain interactions in the stabilization of right-handed twist of $\beta$-sheets. J. Mol. Biol. 168, 389--407.\\

\noindent Chou, K.C., Nemethy, G., Scheraga, H.A., 1990a. Review: Energetics of interactions of regular structural elements in proteins. Acc. Chem. Res. 23, 134--141.\\

\noindent Chou, K. C., Pottle, M., Nemethy, G., Ueda, Y., Scheraga, H. A., 1982. Structure of beta-sheets: Origin of the right-handed twist and of the increased stability of antiparallel over parallel sheets. J. Mol. Biol. 162, 89--112.\\

\noindent Chou, K. C., Scheraga, H. A., 1982. Origin of the right-handed twist of beta-sheets of poly-L-valine chains. Proc. Natl. Acad. Sci. USA 79, 7047--7051.\\

\noindent Chou, K.C., Shen, H.B., 2009. Review: recent advances in developing web-servers for predicting protein attributes. Nat. Sci. 2, 63--92 (http://www.scirp.org/journal/NS/).\\

\noindent Chou, K.C., Wei, D.Q., Zhong, W.Z., 2003. Binding mechanism of coronavirus main proteinase with ligands and its implication to drug design against SARS. (Erratum: ibid., 2003, Vol.310, 675). Biochem. Biophys. Res. Comm. 308, 148--151.\\

\noindent Chou, K.C., Wu, Z.C., Xiao, X., 2011. iLoc-Euk: a multi-label classifier for predicting the subcellular localization of singleplex and multiplex eukaryotic proteins. PLoS ONE 6, e18258.\\

\noindent Chou, K.C., and Zhang, C.T., 1995. Review: Prediction of protein structural classes. Crit. Rev. Biochem. Mol. Biol. 30, 275--349.\\

\noindent Ding, H., Luo, L., Lin, H., 2009. Prediction of cell wall lytic enzymes using Chou's amphiphilic pseudo amino acid composition. Protein Peptide Lett. 16, 351--355.\\

\noindent Du, Q.S., Sun, H., Chou, K.C., 2007. Inhibitor design for SARS coronavirus main protease based on ``distorted key theory". Med. Chem. 3, 1--6.\\

\noindent Du, Q.S., Huang, R.B., Wang, S.Q., Chou, K.C., 2010. Designing inhibitors of M2 proton channel against H1N1 swine influenza virus. PLoS ONE 5, e9388.\\

\noindent Gao, D.Y., 2000. Duality Principles in Nonconvex Systems: Theory, Methods and Applications. Kluwer Academic Publishers, Dordrecht/Boston/London, xviii+454pp.\\

\noindent Gao, D.Y., Ruan, N., Pardalos, P.M., 2010. Canonical dual solutions to sum of fourth-order polynomials minimization problems with applications to sensor network localization, in Sensors: Theory, Algorithms and Applications, P.M. Pardalos, Y.Y. Ye, V. Boginski, and C. Commander (eds). Springer.\\

\noindent Gao, D.Y., N. Ruan, Sherali, H., 2009. Solutions and optimality criteria for nonconvex constrained global optimization problems with connections between canonical and Lagrangian duality. J. Glob. Optim. 45(3), 473--97.\\

\noindent Gao, D.Y., Wu, C.Z., 2012. On the triality theory in global optimization (I) unconstrained problems, arXiv:1104.2970v1, to appear in J. Glob. Optim.:\\ 
http://arxiv.org/PS\_cache/arxiv/pdf/1104/1104.2970v1.pdf\\

\noindent Gong, K., Li, L., Wang, J.F., Cheng, F., Wei, D.Q., Chou, K.C., 2009. Binding mechanism of H5N1 influenza virus neuraminidase with ligands and its implication for drug design. Med. Chem. 5, 242--249.\\

\noindent Griffith, J.S., 1967. Self-replication and scrapie. Nature 215, 1043–-4.\\

\noindent Grosso, A., Locatelli, M., Schoen, F., 2009. Solving molecular distance geometry problems by global optimization algorithms. Comput. Optim. Appl. 43, 23-–37.\\

\noindent Hayat, M., and Khan, A., 2011. Predicting membrane protein types by fusing composite protein sequence features into pseudo amino acid composition. J. Theor. Biol. 271, 10--17.\\

\noindent Holscher, C., Delius, H., Burkle, A., 1998. Overexpression of nonconvertible PrP$^C$ delta114-121 in scrapie-infected mouse neuroblastoma cells leads to trans-dominant inhibition of wild-type PrP$^{Sc}$ accumulation. J. Virol. 72, 1153--9.\\

\noindent Humphrey, W., Dalke, A., Schulten, K., 1996. VMD—visual molecular dynamics. J. Mol. Graph. 14, 33-–38.\\

\noindent Jobling, M.F., Huang, X., Stewart, L.R., Barnham, K.J., Curtain, C., Volitakis, I., Perugini, M., White, A.R., Cherny, R.A., Masters, C.L., Barrow, C.J., Collins, S.J., Bush, A.I., Cappai, R., 2001. Copper and zinc binding modulates the aggregation and neurotoxic properties of the prion peptide PrP 106--126. Biochem. 40, 8073--84.\\

\noindent Jobling, M.F., Stewart, L.R., White, A.R., McLean, C., Friedhuber, A., Maher, F., Beyreuther, K., Masters, C.L., Barrow, C.J., Collins, S.J., Cappai, R., 1999. The hydrophobic core sequence modulates the neurotoxic and secondary structure properties of the prion peptide 106--126. J. Neurochem. 73, 1557--65.\\

\noindent Kandaswamy, K.K., Chou, K.C., Martinetz, T., Moller, S., Suganthan, P.N., Sridharan, S., Pugalenthi, G., 2011. AFP-Pred: A random forest approach for predicting antifreeze proteins from sequence-derived properties. J. Theor. Biol. 270, 56--62.\\

\noindent Kuwata, K., Matumoto, T., Cheng, H., Nagayama, K., James, T.L., Roder, H., 2003. NMR-detected hydrogen exchange and molecular dynamics simulations provide structural insight into fibril formation of prion protein fragment 106--126. Proc. Natl. Acad. Sci. USA 100, 14790--5.\\

\noindent Liao, Q.H., Gao, Q.Z., Wei, J., Chou, K.C., 2011. Docking and molecular dynamics study on the inhibitory activity of novel inhibitors on epidermal growth factor receptor (EGFR). Med. Chem. 7, 24--31.\\

\noindent Lin, H., Ding, H., 2011. Predicting ion channels and their types by the dipeptide mode of pseudo amino acid composition. J. Theor. Biol. 269, 64--69.\\

\noindent Mohabatkar, H., 2010. Prediction of cyclin proteins using Chou's pseudo amino acid composition. Protein Peptide Lett. 17, 1207--1214.\\

\noindent More, J.J., Wu, Z.J., 1997. Global continuation for distance geometry problems. SIAM J. Optim. 7, 814-–836.\\

\noindent Norstrom, E.M., Mastrianni, J.A., 2005. The AGAAAAGA palindrome in PrP is required to generate a productive PrP$^{Sc}$--PrP$^C$ complex that leads to prion propagation. J. Biol. Chem. 280, 27236--43.\\

\noindent Oxenoid, K., Chou, J. J., 2005. The structure of phospholamban pentamer reveals a channel-like architecture in membranes. Proc. Natl. Acad. Sci. USA 102, 10870--10875.\\

\noindent Pielak, R. M., Chou, J. J., 2010. Solution NMR structure of the V27A drug resistant mutant of influenza A M2 channel. Biochem. Biophys. Res. Commun. 401, 58--63.\\

\noindent Pielak, R. M., Chou, J. J., 2011. Influenza M2 proton channels. Biochim. Biophys. Acta 1808, 522--529.\\

\noindent Pielak, R. M., Jason, R., Schnell, J. R., Chou, J. J., 2009. Mechanism of drug inhibition and drug resistance of influenza A M2 channel. Proc. Natl. Acad. Sci. USA 106, 7379-7384.\\

\noindent Sawaya, M.R., Sambashivan, S., Nelson, R., Ivanova, M.I., Sievers, S.A., Apostol, M.I., Thompson, M.J., Balbirnie, M., Wiltzius, J.J., McFarlane, H.T., Madsen, A., Riekel, C., Eisenberg, D., 2007. Atomic structures of amyloid cross-beta spines reveal varied steric zippers. Nature 447, 453--7.\\

\noindent Schnell, J. R. Chou, J. J., 2008. Structure and mechanism of the M2 proton channel of influenza A virus. Nature 451, 591--595.\\

\noindent Sun, J., Zhang, J.P., 2001. Global convergence of conjugate gradient methods without line search. Ann. Oper. Res. 103, 161–-173.\\

\noindent Tsai, H.H.G., 2005. Understanding the biophysical mechanisms of protein folding, misfolding, and aggregation at molecular level (in Chinese). Chem. (The Chinese Chem. Soc. of Taipei) 63, 601-–12.\\

\noindent Wagoner, V., 2010. Computer Simulation Studies of Self-Assembly of Fibril-forming Peptides with an Intermediate Resolution Protein Model, PhD thesis, North Carolina State University, Raleigh, North Carolina.\\

\noindent Wang, J.F., Chou, K.C., 2010. Insights from studying the mutation-induced allostery in the M2 proton channel by molecular dynamics. Protein Eng. Des. Sel. 23, 663--666.\\

\noindent Wang, J., Pielak, R. M., McClintock, M. A., Chou, J. J., 2009. Solution structure and functional analysis of the influenza B proton channel. Nat. Struct. Mol. Biol. 16, 1267--1271.\\

\noindent Wang, J.F., Wei, D.Q., Li, L., Chou, K.C., 2008. Review: Drug candidates from traditional Chinese medicines. Curr. Top. Med. Chem. 8, 1656--65.\\

\noindent Wang, J.F., Yan, J.Y., Wei, D.Q., Chou, K.C., 2009. Binding of CYP2C9 with diverse drugs and its implications for metabolic mechanism. Med. Chem. 5, 263--270.\\

\noindent Wegner, C., Romer, A., Schmalzbauer, R., Lorenz, H., Windl, O., Kretzschmar, H.A., 2002. Mutant prion protein acquires resistance to protease in mouse neuroblastoma cells. J. Gen. Virol. 83, 1237--45.\\

\noindent Wei, H., Wang, C.H., Du, Q.S., Meng, J., Chou, K.C., 2009. Investigation into adamantane-based M2 inhibitors with FB-QSAR. Med. Chem. 5, 305--317.\\

\noindent Wei, D.Q., Sirois, S., Du, Q.S., Arias, H.R., Chou, K.C., 2005. Theoretical studies of Alzheimer's disease drug candidate [(2,4-dimethoxy) benzylidene]-anabaseine dihydrochloride (GTS-21) and its derivatives. Biochem. Biophys. Res. Commun. 338, 1059--64.\\

\noindent Zeng, Y.H., Guo, Y.Z., Xiao, R.Q., Yang, L., Yu, L.Z., Li, M.L., 2009. Using the augmented Chou's pseudo amino acid composition for predicting protein submitochondria locations based on auto covariance approach. J. Theor. Biol. 259, 366--372.\\

\noindent Zhang, J.P., 2009. Studies on the structural stability of rabbit prion probed by molecular dynamics simulations. J. Biomol. Struct. Dyn. 27, 159–-62.\\

\noindent Zhang, J.P., 2011. Optimal molecular structures of prion AGAAAAGA amyloid fibrils formatted by simulated annealing. J. Mol. Model. 17, 173–-9 (Crystallography Time 3(1), January 2011 \& and VirticalNews 2011 FEB 1:\\ 
http://www.rigakumsc.com/downloads/newsletter/LifeSciencesV03N01.html,\\
http://technology.verticalnews.com/articles/4831360.html).\\

\noindent Zhang, J.P., Sun, J., Wu, C.Z., 2011. Optimal atomic-resolution structures of prion AGAAAAGA amyloid fibrils. J. Theor. Biol. 279(1), 17–-28 (Nuclear Energy Research Today Volume 7 Issue 4, April 2011, arXiv:1012.2504v6:\\ http://nuclearenergy.researchtoday.net/archive/7/4/4311.htm)\\

\noindent Zheng, J., Ma, B.Y., Tsai, C.J., Nussinov, R., 2006. Structural stability and dynamics of an amyloid-forming peptide GNNQQNY from the yeast prion Sup-35. Biophy. J. 91, 824-–33.\\

\noindent Zhou, X.B., Chen, C., Li, Z.C., Zou, X.Y., 2007. Using Chou's amphiphilic pseudo-amino acid composition and support vector machine for prediction of enzyme subfamily classes. J. Theor. Biol. 248, 546--551.\\

\noindent Zou, Z.H., Bird, R.H., Schnabel, R.B., 1997. A stochastic/perturbation global optimization algorithm for distance geometry problems. J. Glob. Optim. 11, 91–-105.\\
\end{document}